# Electronic Transport Imaging in a Multiwire SnO$_2$ ChemFET Device

Sergei V. Kalinin,[*] J. Shin, S. Jesse, D. Geohegan, and A.P. Baddorf

Condensed Matter Sciences Division, Oak Ridge National Laboratory, Oak Ridge, TN 37831

Y. Lilach[♣], M. Moskovits and A. Kolmakov[†]

Dept. of Chemistry and Biochemistry, University of California Santa Barbara, CA 93106
[♣]Pacific Northwest National Laboratory, Richland, WA 99352
[†]Physics Department, Southern Illinois University Carbondale, IL 62901

## ABSTRACT

The electronic transport and the sensing performance of an individual SnO$_2$ crossed-nanowires device in a three-terminal field effect configuration were investigated using a combination of macroscopic transport measurements and Scanning Surface Potential Microscopy (SSPM). The structure of the device was determined using both Scanning Electron- and Atomic Force Microscopy data. The SSPM images of two crossed 1D nanostructures, simulating a prototypical nanowire network sensors, exhibit large dc potential drops at the crossed-wire junction and at the contacts, identifying them as the primary electroactive elements in the circuit. The gas sensitivity of this device was comparable to those of sensors formed by individual homogeneous nanostructures of similar dimensions. Under ambient conditions, the DC transport measurements were found to be strongly affected by field-induced surface charges on the nanostructure and the gate oxide. These charges result in a memory effect in transport measurements and charge dynamics which are visualized by SSPM. Finally, scanning probe microscopy is used to measure the current-voltage characteristics of individual active circuit elements, paving the way to a detailed understanding of chemical functionality at the level of an individual electroactive element in an individual nanowire.

---

[*] sergei2@ornl.gov
[†] akolmakov@physics.siu.edu

I. **INTRODUCTION**

The unique transport properties of one-dimensional structures such as oxide and semiconductor nanowires and carbon nanotubes make them promising materials for chemical and biological sensors, photovoltaic and nanoelectronic devices and other applications.[1,2,3,4] The relatively large Debye length of moderately-doped oxides, as compared to the diameter of the nanowires increases the likelihood that even a low density of chemical or biological agents on the surface can result in the depletion (or accumulation) of electrons (holes) in the nanostructure with an associated change in resistivity. Combined with the large surface to volume ratio inherent in one-dimensional nanostructures, this provides the basis for superior sensor function by the oxide nanowire device. Currently two major strategies are employed for fabricating planar chemical sensors based on 1-D nanostructures configured as simple conductors (chemiresistors) or field effect transistors (chemFET): the nanostructure can be wired up as an active sensing element (Fig.1a), or as a percolating network of many nanostructures. In the former, the interaction of donor or acceptor molecules with the nanostructure surface wherein charge is transferred between the surface and adsorbate, alters the equilibrium concentration of free carriers inside the bulk of the nanostructure. Thus, the entire surface of the nanostructure acts as an electroactive element, effectively transducing the surface chemical process into an electrical signal. Despite its obvious advantages, the multi-step fabrication-alignment-wiring procedures required to access individual nanostructures makes this approach experimentally challenging. An alternative is to deposit relatively widely spaced electrodes onto mats of 1-D nanostructures. This reduces significantly the fabrication difficulties and results in device architectures amenable to mass production. However, the device's operation then involves both transport within individual nanowires and across junctions between them, since most nanowires would have no direct electrical contact with the electrodes. The role of junctions (or "necks") between 1-D nanostructures in such nanowire networks can be the dominant factor governing the overall sensor response. This is especially the case for nanostructures with diameters exceeding the Debye length (such as nanostructured thin-films composed of interconnecting nano- and mesoscopic particles) where the conduction channel includes the additional barriers due to the two depletion layers at the nanowires' surface (Fig.1 b). The change in conductance due to the barriers can exceed those induced by gas adsorption, since tunneling or thermionic emission through the depletion region has much dramatic dependence on barrier width than the simple power law governing transport through the nanowire due to adsorption.

The importance of spatially resolved transport measurements on 1D nanostructures has recently been illustrated in carbon nanotubes by scanning probe microscopy (SPM),[5,6,7,8] which clearly indicated that their transport properties are controlled by a small number of active sites such as atomic defects or nanotube to metal contacts. These studies show two key points: (1) nanowires cannot be considered to be scaled-down versions of macroscopic wires and; (2) their electrical behavior can not be deduced from morphology alone. Transport measurements at the nanometer level can be carried out using current sensitive SPM in the contact mode for which the resolution is determined by the tip-surface contact.[9,10] However, current-based SPM techniques are extremely sensitive to surface cleanliness. Even the minute contaminants in the tip-surface junction or the presence of an intrinsic depletion region at the oxide surface will result in resistances that are comparable to the input impedance of the measuring circuit. An alternative is to use non-contact force-based SPM techniques, which are significantly less sensitive to surface depletion or to contamination yet allow imaging at very small tip-sample forces, which is particularly helpful in studying fragile structures. However, because the resolution of an ambient electrostatic SPM (~ 100 – 300 nm)[11] is comparable to the defect spacing (~100 - 300 nm) quantitative local I-V measurements have, until now, not been carried out.

$SnO_2$ based nanowires and nanobelts are promising new structures as catalysts and gas sensors.[12,13,14] Here we use SPM-based potential measurements to characterize transport in a $SnO_2$ 1-D nanostructure sensor, which consists of two crossing electroactive elements forming a primitive "network". We determine the geometry of the two nanowires, and elucidate the transport from their individual I-V characteristics. The response of this "network sensor" towards minute oxygen exposure is tested and compared with performance of individual homogeneous nanobelt chemiresistors with similar dimensions.

## II. EXPERIMENTAL

$SnO_2$ 1-D nanostructures were synthesized as described in Refs. [15,16]. Briefly, single crystal $SnO_2$ (rutile) nanowires and nanobelts were vapor grown in a tube furnace by thermal evaporation of SnO at 1000 °C into an Ar carrier gas (50 sccm, 200 Torr) containing traces of oxygen. The structure, stoichiometry and morphology of the resultant nanostructures were verified with SEM, HRTEM, Raman spectroscopy and XRD. The nanowires and nanobelts, collected from the ceramic crucible, were placed on the oxide side of a $Si/SiO_2$ (300 nm) wafer. Electrical contacts to the nanostructure were vapor deposited as Ti (20 nm)/Au (200 nm) micro-pads which acted as the source and drain electrodes. To avoid contamination of the nanostructures, the electrodes were deposited in vacuum through a shadow mask. No wet processes were used such as those in resist-based lithography in order to avoid the electrical complications common with contact fashioned by photo- or e-beam lithography. The (p-doped) Si substrate was used as a conducting back gate electrode. The device was wire bonded and placed onto a custom-designed chip holder for imaging with SPM under applied bias.

The gas sensing measurements of the same nanostructure were conducted in a variable-pressure probe station. Before the measurements were carried out, the sample was cleaned and annealed in vacuum ~$10^{-5}$ Torr for approximately an hour at T = 200 °C. Oxygen pulses of 1-2 x$10^{-3}$ Torr were introduced into the chamber using pulsed leak valves. The resultant changes in source-drain current were measured as a function of time at a bias $V_{DS}$ =2 V.

Spatially resolved AFM and SPM-based dc transport measurements were performed on a commercial SPM system (Veeco MultiMode NS-IIIA) equipped with a custom-built sample holder, which allows *in-situ* biasing of the nanowire sample. The sample was connected in series with 100 kOhm current-limiting resistors as shown in Fig. 1c,d. The biases on electrodes and back-gate were controlled independently by voltage function generators (DS 345 and 340, Stanford Research Instruments). Measurements were performed in the Scanning Surface Potential Microscopy (SSPM)[17,18,19] mode using Pt coated tips (NCSC-12 F, Micromasch, $l \approx 250$ μm, resonant frequency ~ 41 kHz) with typical lift heights of 200 nm. The current is imposed laterally across the surface using macroscopic electrodes similar to those used for 4-probe resistance measurements. Here, the SPM tip is used as a moving voltage-sensing electrode, providing a spatially-resolved dc potential distribution image along the nanowire.

### III. RESULTS AND DISCUSSION

## A. The morphology of the sample

The surface topography of a sample containing the simplest nanowire network building block is shown in Fig. 2a. The sample consists of two crossed nanowires ~85 nm and ~ 25 nm in height. The lateral size cannot be reliably determined using AFM due to the convolution with the tip shape. To unambiguously establish the geometric structure of the network, the sample was subsequently examined by SEM, which provides more reliable lateral dimensions. The SEM images in Fig. 2b show that while for the smaller nanowire the lateral size is comparable with the ~25 nm height measured by AFM, for the large nanowire the lateral dimension is ~650 nm, significantly larger than the apparent height of 85 nm determined by AFM. Combining the AFM and SEM measurements, the individual elements forming the network are determined to be a 85 nm x 650 nm nanobelt positioned over a ~ 25 nm diameter nanowire. The nanobelt makes contact with both electrodes, whereas the smaller nanowire makes contact with the bottom electrode only.

## B. Gas sensing measurements

The chemical sensing of the crossed nanostructure system was tested and its performance was compared with the "ideal" nanowire sensor based on a single homogeneous nanobelt of the similar dimensions, grown and wired in the same run and therefore exposed to the same treatment as a structure of interest. One therefore expects the observed differences in the performance of these two structures to be primarily determined by their morphological differences rather than electrode or substrate effects. The measured source-drain ($I_{DS}$) current at $V_{DS}$ =2 V for both nanostructures was found to depend critically on the composition of the ambient gas. In Figure 3(a) the responses to oxygen and hydrogen pulses of ~$10^{-3}$ Torr of a single nanobelt sensor are shown. The "network" device exhibited the comparable behavior for both oxidizing and reducing agents, which is shown in details in the Fig. 3(b) for a single oxygen pulse. Under vacuum (and after approximately an hour's annealing at 200 °C) both nanostructures behaved as fairly good conductors, with comparable conductivities ($\sigma_{(VG=0)}$ = 90.5 $(\Omega\ cm)^{-1}$ at 200 C°). Upon admitting $5\cdot10^{-4}$ Torr of oxygen (see Fig.3(b)), for~ 200 s the nanostructures' conductivity drops dramatically to $\sigma_{(VG=0)}$ = 27 $(\Omega\ cm)^{-1}$ for the crossed nanstructured system and 9 $(\Omega\ cm)^{-1}$ for the single nanowire. The origin of this phenomenon is well understood for thin and thick film sensors based on $SnO_2$.[20] For $SnO_2$ nanostructures at ~200 °C and at zero gate potential, the material's high conductance in vacuum results from the presence of shallow, i.e. totally ionized donor states consisting of a high density of oxygen vacancies. The latter also render the oxide an *n*-type semiconductor. Under these conditions the Fermi level is just below the conduction band edge (see Fig. 1 b). Exposure to oxygen saturates the surface vacancies, drawing electrons from the bulk and localizing them on the chemisorbed oxygen molecules. Due to its high surface-to-bulk ratio, the small number of electrons in the nanostructure and the facile access of the bulk electrons for surface processes, this electron depletion results in a significant drop in conductance even for such low concentration of oxidant. For reducing gases like hydrogen the sensing-transduction mechanism is reverse (see recent review [Ref. 20] and references therein). Two important differences between the two nanostructures were noticed. (i) Unlike the "network", the single nanowire is characterized by ohmic current-voltage behavior at 200 °C both in vacuum and at ca $10^{-3}$ Torr of oxygen. (ii) The low stability of the $I_{DS}$ current is seen in curve B (Fig.3(b)) for the crossed-nanowire structure.

These observations can be rationalized assuming that the crossed-nanowire junction does indeed modify the conduction and sensing performance of the entire device and transport

SSPM measurements illustrated below corroborate with this observation. In a homogeneous ideal nanobelt with thickness (T ~ $2\lambda_D$) oxygen chemisorption induces a change in the electron density uniformly throughout the entire length of the nanostructure, constricting the effective diameter of the conduction channel. The ohmic current–voltage characteristics also indicate that at 200 C°electrons readily overcome Schottky barriers, which arise primarily at the contacts. Since the amount of adsorbed species is statistically large, the resultant current is a smooth function of adsorbate coverage. By contrast, the observed current instability and the non-linearity of the current-voltage characteristics suggest that the mechanism described above is not the dominant one for the crossed-nanostructures, in which the conducting channel has at least one additional resistance, namely the junction barrier in series with the nanobelt, as well as a second nanowire and the Schottky barriers at the contacts, all of which are sensitive to oxygen. The electron transport and the sensing response of the crossed structure will therefore depend on the electroactive element with the largest resistance. We ascribe the observed I-V nonlinearity and the increased current noise to the formation of a gas sensitive electroactive element at the nanowire junction. In spite of the small lateral size of this electroactive element, the sensitivity of the device is comparable to the single nanobelt device. Overall, the sensitivity is quite high. Operating these devices in an inert gas at normal pressures, the sensitivity of both toward oxygen and hydrogen would be sufficient for detection at the ppb level.

**C. Imaging of the electron transport in the device**

The local transport was addressed by room temperature SSPM studies on the same device. A series of surface potential images measured at different bias voltages is shown in Fig. 4. Biasing the device results in potential drops at the top contact, the junction between the two nanostructures, and the bottom contact, while the potential is nearly uniform along the individual nanobelt segments. These data are shown quantitatively in Fig. 5a in which the absolute potential distribution across the nanobelt is plotted as a function of device bias. The potential distribution inside individual segments is virtually constant and significant potential drops occur only at these electroactive sites.

The images in Fig. 4 shows that, when a negative bias is applied to the top electrode, both segments of the nanobelt are biased (i.e. both segments are bright in Fig. 4d), while with positive bias only one segment is at an appreciable potential (i.e. only the top segment in Fig. 4c is bright). Hence, applying a +6 V bias to the top electrode is not equivalent to applying a -6V bias to the lower electrode, since in both cases the back (gate) electrode is grounded, resulting in different gate conditions for the nanowire in the two instances. Comparing Fig. 4c,d, shows that the transport properties of some of the electroactive elements in the network are asymmetric. This behavior is further illustrated in Fig. 5b,c which displays the potential difference along the nanowire (referenced to the surrounding material) after the linear ohmic contribution is subtracted. The potential difference between the two electrodes is close to the electrostatic potential of the nanostructure. With a negative electrode bias, the differential potential profiles measured with the top- and bottom electrode biased are symmetric (Fig. 5 c), while with a positive bias (Fig. 5 b) there is strong asymmetry between potential profiles, indicating the presence of at least one rectifying element in the circuit. In all cases, the potential distribution is almost uniform (black curve) when both electrodes are biased.

In all cases, the width of the potential step is several microns, significantly larger than the resolution in the SSPM images as determined by using calibration standards.[11,21] One significant factor that can affect dc potential measurements by SSPM is the presence of mobile surface charges.[22,23]

**D. The effects of ambient: the formation mobile surface charges and "memory" effects**

The "parasitic" charging is observed in our sample as lateral spreading (with characteristic time constant in the order of 20-30 minutes) of the potential distribution under continuous bias and the retention of an electric field after the bias is turned off, as illustrated in Fig. 6. Shown in Fig. 6a,b,c is the surface potential image of biased device after 10 min, 20 min and 1 h under bias. Notice the gradual spreading of the potential due to the surface charge mobility. Reverse behavior is observed after the bias is turned off, as shown in Fig. 6d,e,f. Here, the dark halo can be clearly seen after the negative bias was turned off, indicative of the presence of mobile charges. After the relaxation for a week, the halo disappears. Notice that in this case the charge is injected from the electrode and the sign of the charge is that of the bias, i.e. the surface and the electrode are coupled resistively.

This surface charge effect can be also observed in the macroscopic transport measurements as a response to the back gate bias. Shown in Fig. 7a is a transient response of the nanowire to abrupt change of the gate potential from -8.7 V to 0. Here, application of the negative potential to the back gate results in the positive charge accumulation on the oxide surface. After the gate is switched off, the positive surface charge gates the nanowire similarly to positive back gate. Notice that the relaxation time of these charges is of order of hours, similarly to SPM observations. In comparison, in the blank experiment when nanowire is not present between the contact pads, no long term response is observed. This behavior is also reflected in the time-dependence of I-V curves taken sequentially after the sudden change of back gate bias, as illustrated in Fig. 7b. These observations illustrate that the mobile charge effect is also observed in the back gate experiments; however, in this case the sign of the screening charge is opposite to back gate bias, as can be expected given that backgate and surface are coupled capacitively.

These surface charge dynamics can strongly affect the chemical and biological sensing performance of oxide-nanowire-based devices operating in real world wet or humid environment. The mobile charge effect is the primary limitation of ambient dc transport measurements by SSPM and can be avoided by carrying out high-frequency transport measurements with, for example, Scanning Impedance Microscopy.[24,25] However, applying an AC bias through a low-conductivity material leads to the formation of an alternative capacitive current path through the back gate electrode, restricting the utility of such measurements in samples with large conductive pads such as those used in this work.

**E. The nature of the local particularities in electron transport**

The data presented in Figs. 4 and 5 can be quantified further to yield transport properties of the individual electroactive elements. Potential images were obtained for lateral bias values between the bottom and top electrodes ranging from -8 to 8 V in 0.5 V steps. To minimize the effect of mobile charges, the polarity of the bias was alternated between consecutive images. Fig. 5d shows the potential drop at the bottom electrode, the nanowire junction and the top electrode, corresponding to the locally-determined potential drops across the individual electroactive elements as a function of the applied lateral bias. Simultaneously, the total current across the system was determined as a function of the lateral bias. Assuming that SSPM provides the *precise* value of the potential drop, combining the SSPM and macroscopic I-V data allows the I-V curve of the individual electroactive elements to be determined, and, ideally, its response to *local* external chemical and biological stimuli, paving the way to a detailed understanding of the operation of these metal oxide nanostructures as sensors.

The potential measured by the SSPM tip is a weighted average of the nanowire potential and the potential of the back electrode, $V_{eff} = (C_{nw}V_{nw} + C_{be}V_{be})/(C_{nw} + C_{be})$, where $C_{nw}$ and $C_{be}$ are the nanowire-tip and bottom electrode-tip capacitance gradients and $V_{nw}$ and $V_{be}$ are nanowire and back electrode potentials.[26] For the nanowire biased on both ends, the measured nanowire potential (~ 4 V for positive bias, -5.5 V for negative bias) is close to the bias potential (6 V and – 6 V respectively) (Fig. 4e,f and Fig. 5b,c). Thus, the error in the potential measurements is less than ~30%. This error is likely dominated by the effect of the mobile charge, which increases the effective nanowire cross-section determined by the SSPM. As a result, potential drops on the electrodes are likely to be overestimated by this order of magnitude, whereas potential drops at the junction are likely underestimated.

The macroscopic I-V curve of the device is asymmetric (Fig. 8 a) and follows an approximate power law $I \sim V^{2.34}$ for both positive and negative biases. The ratio of currents for positive and negative bias polarities is shown in the inset to Fig. 8. Taken in combination with the SPM data, this behavior suggests that for large biases, transport is governed by the rectifying metal-nanowire contacts (there are two segments contacting the bottom electrode and only one nanobelt contacting the top electrode), while at low bias values transport is dominated by the junction in the crossed nanowire junction and by bulk resistances. The potential drop over *individual* electroactive defects measured by SSPM can be combined with the macroscopic I-V measurements to yield the I-V curve of an *individual* defect, as shown in Fig. 8b. In an ambient environment however, a detailed analysis of the local I-V curves is hampered by the effects of adsorption, resulting in significant uncertainty in the values of the measured potential. Unexpectedly, the crossed-nanowire-nanobelt junction acted as a current limiting element even though the nanobelt *per se* is contacted at both ends. This behavior suggests that the smaller nanowire possesses significantly lower resistivity than the nanobelt and shorts out the portion of the nanobelt between the bottom electrode and the nanobelt's center. However, given that under typical ambient conditions the $SnO_2$ exists in a electron-depleted state, this explanation is unlikely. The observed phenomena is therefore more likely attributable either to a structural defect at the nanowire-nanobelt junction if one assumes the resistivity of the two nanostructures are comparable, or to the fact that the smaller nanowire acts as a local gate on the nanobelt. The latter is feasible given the small thickness of the nanobelt and the known propensity for negative ions (hydroxyl groups and chemisorbed oxygen molecules) to accumulate on the surfaces of $SnO_2$ nanostructures under ordinary ambient conditions,[22] thus illustrating an unanticipated role of nanowire junctions in network conductance.

From these curves, we estimate the resistance for top and bottom contact and nanowire junction as $R_1$ = 23 GOhm, $R_3$ = 9.9 GOhm, and $R_2$ = 97 GOhm. For thermionic transport expected for low-mobility material, the relationship between the low bias specific contact resistance and potential barrier height is $R = \dfrac{nk}{qA^*T}\exp\left(\dfrac{q\phi_B}{kT}\right)$, where $q = 1.6\cdot10^{-19}$ C is electron charge, $n$ is ideality factor, $k = 1.38\cdot10^{-23}$ J/K is the Boltzmann constant, $T$ is temperature, $A^*$ = 130 A/(cm$^2$ K$^2$) is the Richardson constant and $\phi_B$ is Schottky potential barrier height. Estimating the cross-section area from the combination of the SEM and AFM data as 5.52 10$^{-10}$ cm$^2$, the effective potential barriers can be estimated as $\phi_B(1) = 0.579$ eV, $\phi_B(3) = 0.557$ eV and $\phi_B(2) = 0.617$ eV. Given that effective conductive region can be much smaller than the nanowire cross-section, this provides an upper estimate on local Schottky barrier height.

## IV  CONCLUSIONS

To summarize, spatially-resolved SSPM potential measurements were combined with macroscopic I-V and gas sensing measurements to determine the electron transport behavior in local electroactive elements in crossed 1D metal-oxide nanostructures. The crossed-nanostructure investigated here may be construed to be a model for a sensor based on networks of quasi-1D nanostructures. In the particular device studied here both SSPM and gas sensing measurements indicate that the junction resulting from crossing the quasi-1D nanostructures dominates the sensing properties of the entire device. A consequence of this observation is that the observed current instabilities inherent in an individual junction can be eliminated in real-world devices by increasing the total number of electroactive elements working in parallel. Excellent sensing performance was recently demonstrated for these devices.[27] For a working device exposed to air with average humidity 50-70%, dc potential measurements are strongly influenced by mobile charges induced on the gate oxide. These charges reduce the spatial resolution and exhibit large relaxation times, producing memory effects. For the $SnO_2$ devices studied here, the properties of metal-nanowire and nanowire-nanowire contacts are found to be rectifying and significantly varying within a device, resulting in macroscopic I-V curves that are asymmetric with respect to interchanging the polarity of the applied bias.

## ACKNOWLEDGEMENTS

Research performed as a Eugene P. Wigner Fellow and staff member at the Oak Ridge National Laboratory, managed by UT-Battelle, LLC, for the U.S. Department of Energy under Contract DE-AC05-00OR22725 (SVK). Support from ORNL Laboratory Research and Development funding is acknowledged (SVK, APB, and DBG). The work at UCSB was supported via AFOSR DURINT grant F49620-01-1-0459, and by funding through the Institute for Collaborative Biotechnologies supported by the US Army.

**Figure captions**

**Figure 1.** (a) Schematic band diagram of an ideal semiconductor 1-D nanostructure joining two highly-conducting electrodes. (b) A conductometric device based on two crossed quasi-1 D nanostructures. Center: the cross-junction barrier resulting from the partial depletion of the near-surface region. The Schottky barriers at the electrodes are also shown. (c) Experimental set-up for carrying out spatially-resolved transport measurements and (d) topographic image of $SnO_2$ nanowire. The lateral size is 50x50 $\mu m^2$ and the vertical scale is 500 nm.

**Figure 2.** (a) Topographic AFM image of the nanostructure and (b) SEM image of the same system. By combining the vertical dimensions determined by AFM and the lateral sizes determined by SEM, the dimensions of the 1D elements were determined as a 650 nm x 85 nm nanobelt and a ~25 nm diameter nanowire.

**Figure 3.** (a) Typical response $I_{DS}$ ($V_{SD}$=2 V) of the individual homogeneous nanobelt to two sequentional ~$10^{-3}$ Torr oxygen (oxidizing agent) pulses followed by two sequential pulses of hydrogen (reducing agent) at 250 °C. (b) The comparison of the response functions of two nanostructures ($I_{DS}$ ($V_{SD}$=2 V), at 200 C°) toward the oxygen pulse. The bold curve (A) corresponds to measurements on a defect-free single nanobelt with ohmic contacts. The narrow line (B) represents measurements for the nanobelt crossed by a nanowire, which also forms Schottky barriers at the contacts. $I_{DS}$ partially recovers in (A) after oxygen exposure, due to the partial evaporation of chemisorbed oxygen from the nanostructure's surface at 200 C°.

**Figure 4.** Surface potential images for biases of 6 V (a,c,e) and -6 V (b, d, f). Voltages were applied to the bottom (a,b), top (c,d) and both (e,f) electrodes. In each case the back gate electrode is grounded. The vertical scale is 10 V.

**Figure 5.** (a) The absolute potential along the nanowire for +6 V applied to bottom and top electrodes. Differential potential profiles for +6 V (b) and -6 V (d) biases. Note the asymmetry between the curves obtained with top and bottom electrodes positively biased. (d) Potential drop on the top contact, nanowire junction and bottom contact as a function of lateral bias.

**Figure 6.** Surface potential on the biased nanowire after (a) 10 min, (b) 20 min and (c) 1 h scanning illustrating the smearing of potential contrast due to the mobile charge effect. (d) surface potential at 10 V negative bias and (e) immediately after the bias is off illustrates the formation of charged negative halo. (f) After a week time, the halo disappears. The scale is 6 V (a,b,c), 10 V (d) and 1 V (e,f).

**Figure 7.** (a) A comparison of the transient response of the wired nanowire (top curve) and empty pads (lower curve) to a sudden change of the gate potential from -8.7 V to 0. The accumulation of the positive mobile charges induced by a negative gate bias gives rise to a positive gating when the bias is off. (b) I-V characteristics of the nanowire (complementary to the top curve in (a)) taken sequentially every few seconds after the sudden change of gate potential.

**Figure 8.** (a) Macroscopic I-V curve of the device. The inset shows the ratio of currents for opposite bias polarities. (b) I-V curves of the top contact (squares), nanowire junction (diamonds) and bottom contact (triangles) from the combination of SSPM and macroscopic transport data. The inset shows the expanded view of the I-V curves.

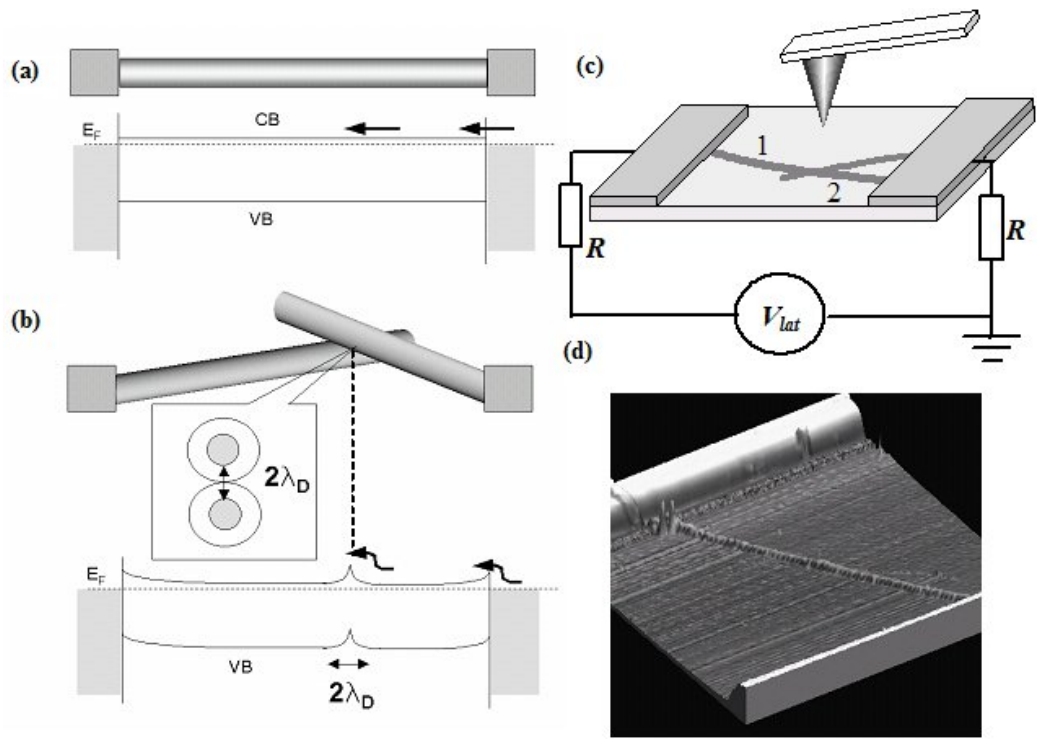

**Figure 1.** S.V. Kalinin, J. Shin, S. Jesse *et al*.

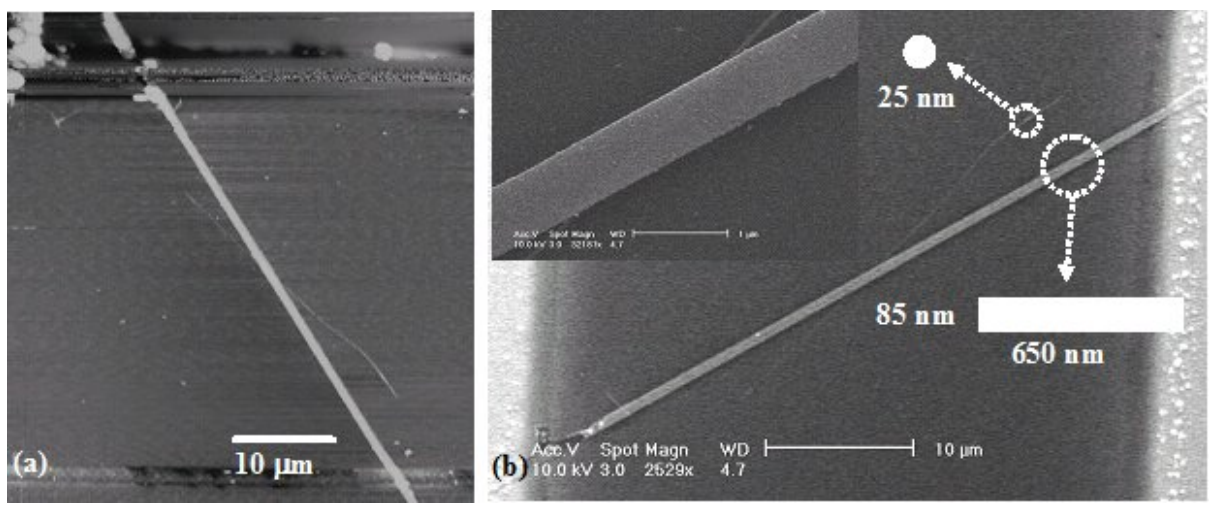

**Figure 2.** S.V. Kalinin, J. Shin, S. Jesse *et al*.

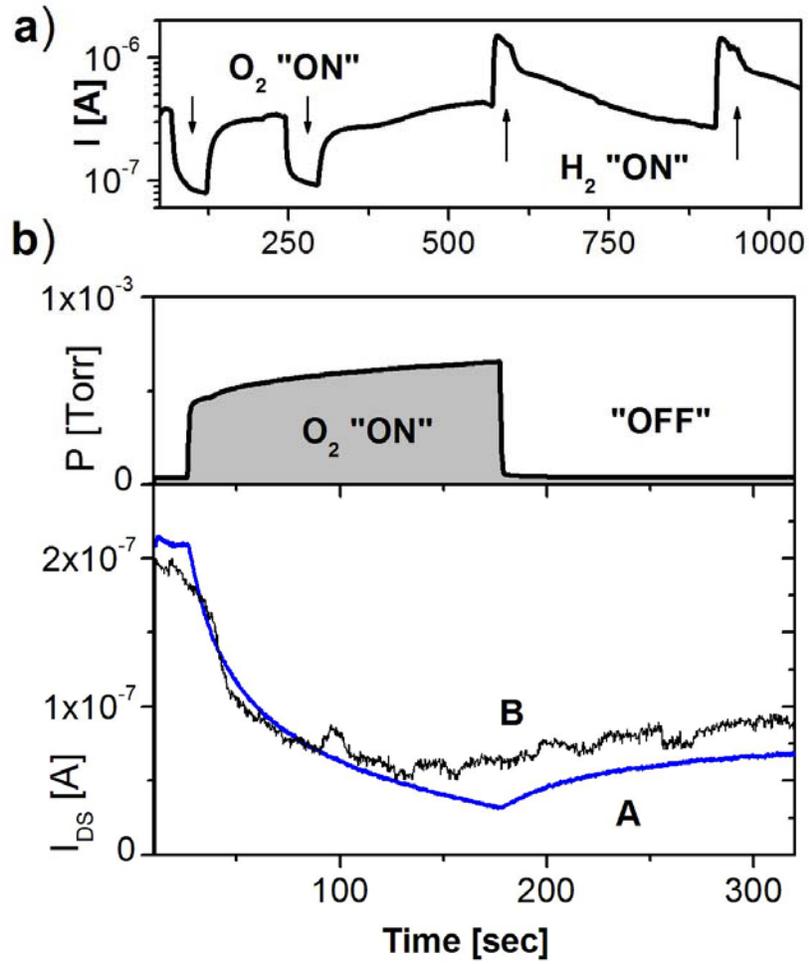

**Figure 3.** S.V. Kalinin, J. Shin, S. Jesse *et al*.

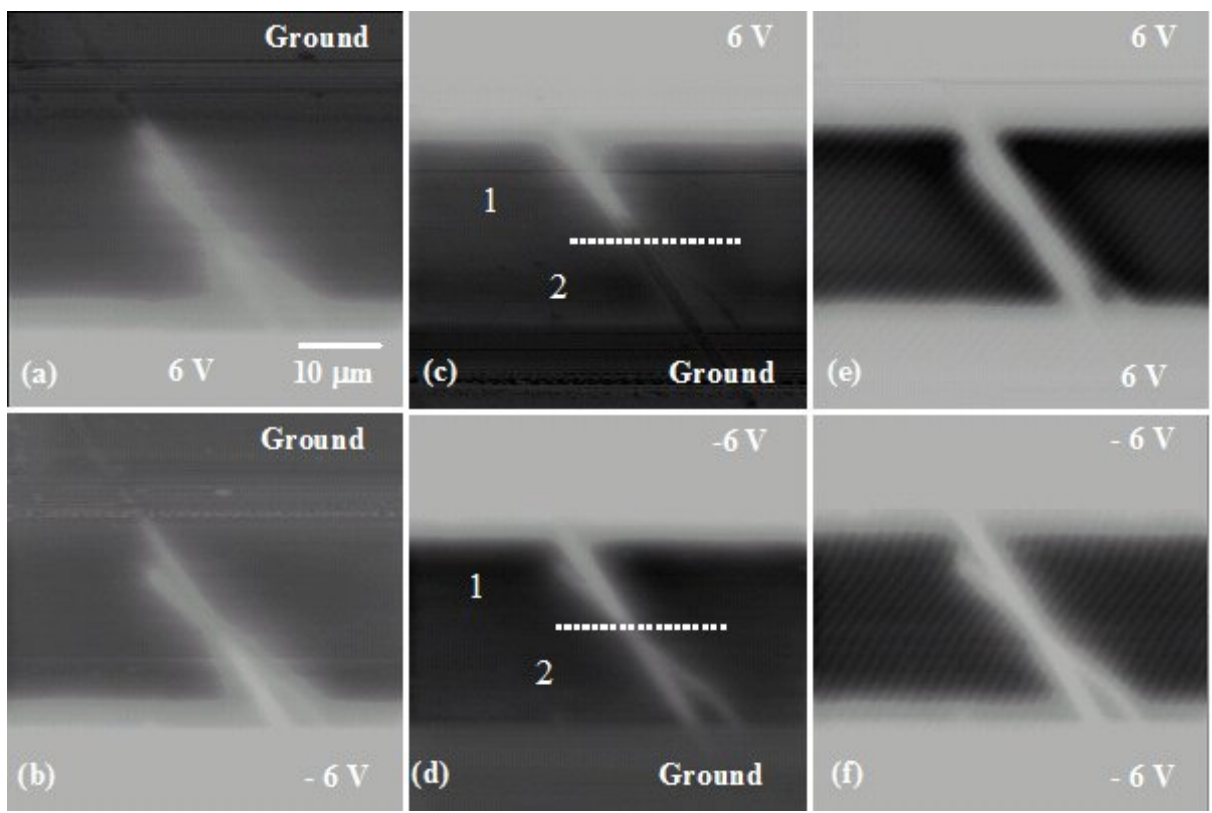

**Figure 4.** S.V. Kalinin, J. Shin, S. Jesse *et al*.

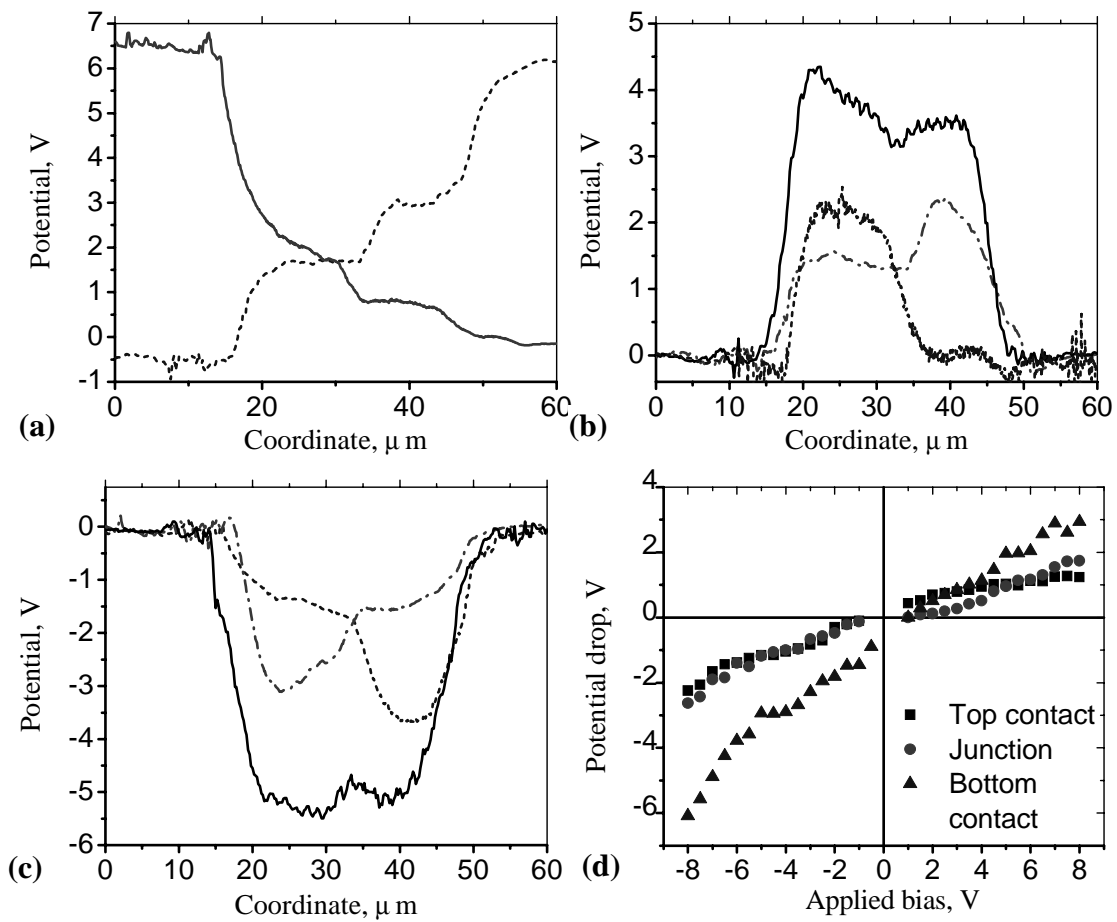

**Figure 5.** S.V. Kalinin, J. Shin, S. Jesse *et al*.

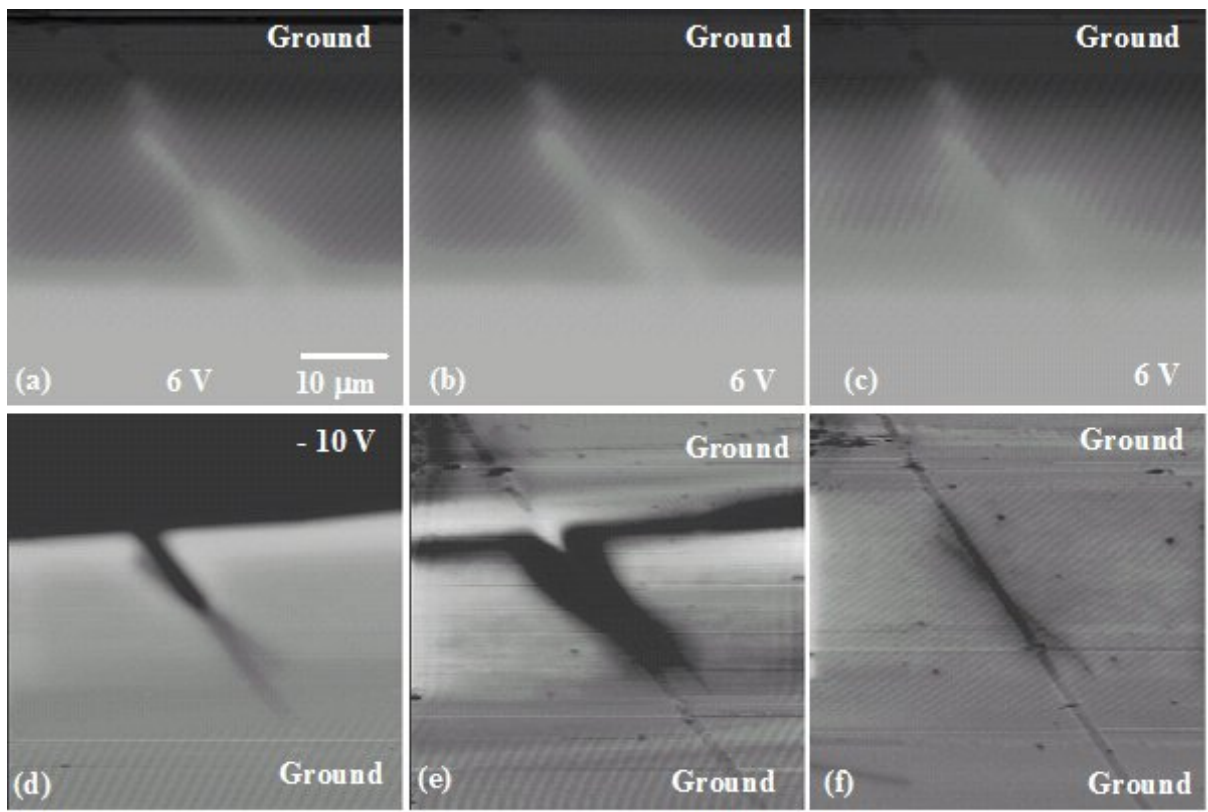

**Figure 6.** S.V. Kalinin, J. Shin, S. Jesse *et al.*

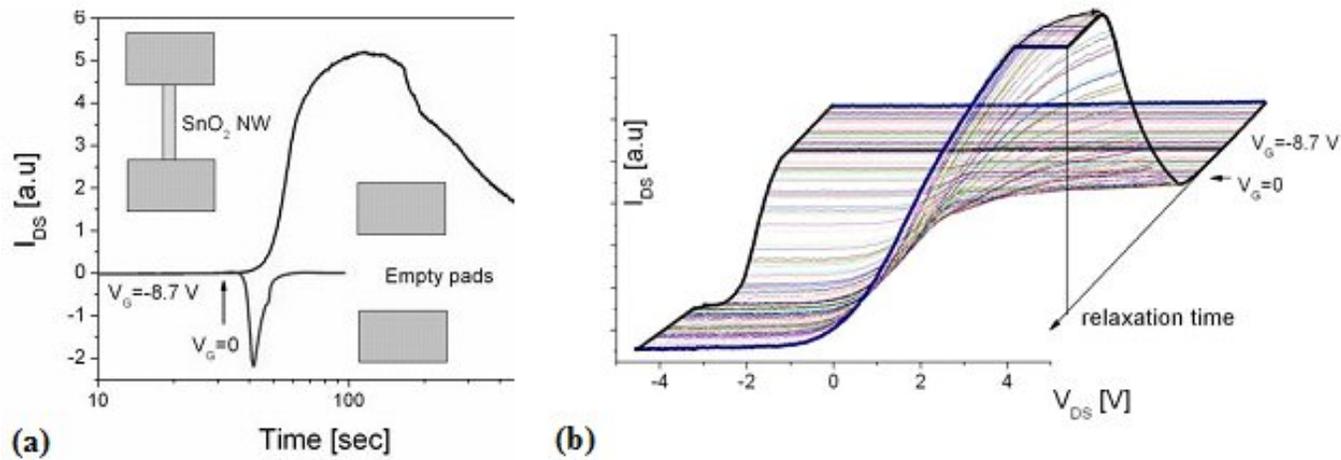

**Figure 7.** S.V. Kalinin, J. Shin, S. Jesse *et al*.
26

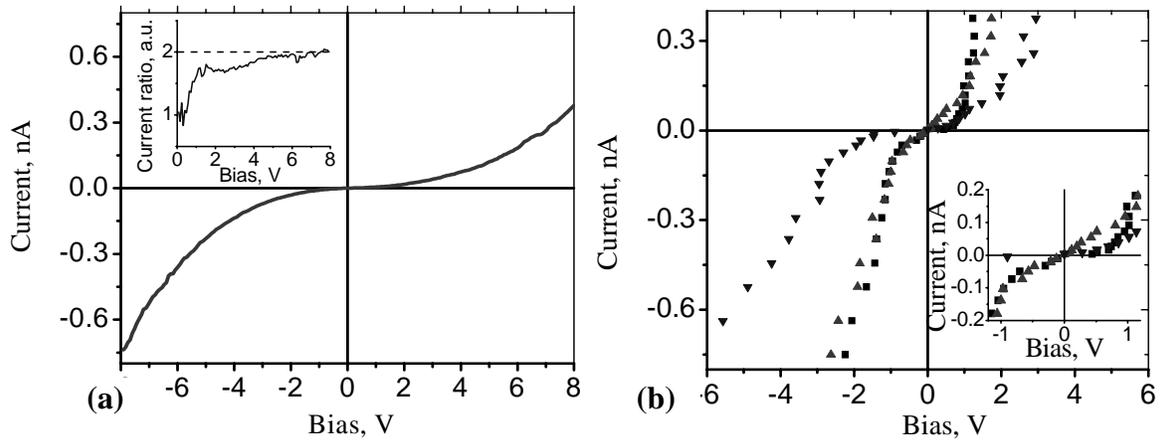

**Figure 8.** S.V. Kalinin, J. Shin, S. Jesse *et al*.